\documentclass[prx,aps,showpacs,twocolumn,preprintnumbers,nofootinbib,amsmath,amssymb,superscriptaddress,longbibliography]{revtex4-2}
\usepackage[T1]{fontenc}
\usepackage[english]{babel}
\usepackage{amsmath,amssymb,amsfonts}
\usepackage{graphicx}
\usepackage[colorlinks=True,linkcolor=red,citecolor=blue,urlcolor=blue]{hyperref}
\usepackage{bookmark}
\usepackage{cancel}
\usepackage{xcolor}
\usepackage{chngcntr}
\usepackage{braket}
\usepackage{nicefrac}
\usepackage{comment}
\usepackage{bm}
\usepackage{bbm}
\usepackage{braket}
\usepackage{slashed}
\usepackage{chemmacros}
\chemsetup{modules=all}
\usepackage[normalem]{ulem}
\usepackage{array}
\newcolumntype{C}{>{$}c<{$}}

\definecolor{macouleur}{RGB}{105,150,150}

\begin{document}
\bibliographystyle{apsrev4-1}
\title{Giant quantum oscillations in thermal transport in low-density metals\\ {\it via} electron absorption of phonons}

\author{B. Bermond}
\email{baptiste.bermond@ens-lyon.fr}
\affiliation{ENSL, CNRS, Laboratoire de Physique, F-69342 Lyon, France}

\author{R. Wawrzy\'{n}czak}
\affiliation{Max Planck Institute for Chemical Physics of Solids, Nöthnitzer Stra{\ss}e 40, 01187 Dresden, Germany}

\author{S. Zherlitsyn}
\affiliation{Hochfeld-Magnetlabor Dresden (HLD-EMFL) and Würzburg-Dresden Cluster of Excellence ct.qmat, Helmholtz-Zentrum Dresden-Rossendorf, 01328 Dresden, Germany}

\author{T. Kotte}
\affiliation{Hochfeld-Magnetlabor Dresden (HLD-EMFL) and Würzburg-Dresden Cluster of Excellence ct.qmat, Helmholtz-Zentrum Dresden-Rossendorf, 01328 Dresden, Germany}

\author{T. Helm}
\affiliation{Hochfeld-Magnetlabor Dresden (HLD-EMFL) and Würzburg-Dresden Cluster of Excellence ct.qmat, Helmholtz-Zentrum Dresden-Rossendorf, 01328 Dresden, Germany}

\author{D. Gorbunov}
\affiliation{Hochfeld-Magnetlabor Dresden (HLD-EMFL) and Würzburg-Dresden Cluster of Excellence ct.qmat, Helmholtz-Zentrum Dresden-Rossendorf, 01328 Dresden, Germany}

\author{G. D. Gu}
\affiliation{Condensed Matter Physics and Materials Science Department, Brookhaven National Laboratory, Upton, NY, USA}

\author{Q. Li}
\affiliation{Condensed Matter Physics and Materials Science Department, Brookhaven National Laboratory, Upton, NY, USA}

\author{F. Janasz}
\affiliation{Department of Engineering, Faculty of Science, Medicine and Technology, University of Luxembourg, Luxembourg}

\author{T. Meng}
\affiliation{Institute for Theoretical Physics and Würzburg-Dresden Cluster of Excellence ct.qmat, Technische Universität Dresden, Dresden, Germany}

\author{F. Menges}
\affiliation{Max Planck Institute for Chemical Physics of Solids, Nöthnitzer Stra{\ss}e 40, 01187 Dresden, Germany}

\author{C. Felser}
\affiliation{Max Planck Institute for Chemical Physics of Solids, Nöthnitzer Stra{\ss}e 40, 01187 Dresden, Germany}

\author{ J. Wosnitza}
\affiliation{Hochfeld-Magnetlabor Dresden (HLD-EMFL) and Würzburg-Dresden Cluster of Excellence ct.qmat, Helmholtz-Zentrum Dresden-Rossendorf, 01328 Dresden, Germany}
\affiliation{Institut für Festkörper-und Materialphysik and Würzburg-Dresden Cluster of Excellence ct.qmat, Technische Universität Dresden, 01062 Dresden, Germany}

\author{Adolfo G. Grushin}
\email{adolfo.grushin@neel.cnrs.fr}
\affiliation{Univ. Grenoble Alpes, CNRS, Grenoble INP, Institut N\'eel, 38000 Grenoble, France}

\author{David Carpentier}
\email{david.carpentier@ens-lyon.fr}
\affiliation{ENSL, CNRS, Laboratoire de Physique, F-69342 Lyon, France}

\author{J. Gooth}
\affiliation{Max Planck Institute for Chemical Physics of Solids, Nöthnitzer Stra{\ss}e 40, 01187 Dresden, Germany}
\affiliation{Physikalisches Institut, Universität Bonn, Nussallee 12, 53115 Bonn, Germany}

\author{S. Galeski}
\email{sgaleski@uni-bonn.de}
\affiliation{Physikalisches Institut, Universität Bonn, Nussallee 12, 53115 Bonn, Germany}
\affiliation{Max Planck Institute for Chemical Physics of Solids, Nöthnitzer Stra{\ss}e 40, 01187 Dresden, Germany}
\affiliation{Hochfeld-Magnetlabor Dresden (HLD-EMFL) and Würzburg-Dresden Cluster of Excellence ct.qmat, Helmholtz-Zentrum Dresden-Rossendorf, 01328 Dresden, Germany}

%%%%%%%%%%%%%%%%%%%%%%%%%%%%%%%%%%%%%%%%%%%%%%%%%%%%%%%%%%%%%%%%%%%%%%%%%%%%%

%\date{\today}
% \date{June 17, 2022}
\begin{abstract}

Oscillations of conductance observed in strong magnetic fields are a striking manifestation of the quantum dynamics of charge carriers in solids. 
The large charge carrier density in typical metals sets the scale of oscillations in both electrical and thermal conductivity, which characterize the Fermi surface. 
%
%, role of electron correlations %and charge carrier Berry phase.
%
% In semimetals, the low electronic density and the dominating phonon contribution to heat transport does not mask the quantization of electron orbits in strong magnetic fields as expected. 
% %
% Several observations of giant quantum oscillations in thermal conductivity of semimetals have challenged this picture. 
%
In semimetals, thermal transport at low-charge carrier density is expected to be phonon dominated, yet several experiments observe giant quantum oscillations in thermal transport. 
%
% : the low electronic density and the dominating phonon contribution to heat transport does not mask the quantization of electron orbits in strong magnetic fields as expected. 
%
This raises the question of whether there is an overarching mechanism leading to sizable oscillations that survives in phonon-dominated semimetals. In this work, we show that such a mechanism exists. It relies on the peculiar phase-space allowed for phonon scattering by electrons when only a few Landau levels are filled. Our measurements on the Dirac semimetal \ch{ZrTe5} support this counter-intuitive mechanism through observation of pronounced thermal quantum oscillations, since they occur in similar magnitude and phase in directions parallel and transverse to the magnetic field. Our phase-space argument applies to all low-density semimetals, topological or not, including graphene and bismuth. Our work illustrates that phonon absorption can be leveraged to reveal degrees of freedom through their imprint on longitudinal thermal transport.
% It illustrates how phonon transport can be engineered as a remarkable probe of properties of various exotic degrees of freedom.
\end{abstract}
\maketitle

\section{Introduction}
Magnetic quantum oscillations are a captivating phenomenon revealing the quantum dynamics of electrons in solids.~\cite{ashcroft2022solid,abrikosov2017fundamentals}. They occur due to the quantization of energy levels of electrons when subjected to an external magnetic field, known as Landau levels~(LLs).
The electronic properties of a solid undergo changes as the number of filled Landau levels varies with the magnetic field, manifesting as oscillations in various physical observables.
Among those, the Shubnikov-de Haas oscillations of electrical resistivity~\cite{Schubnikow:1930a,Schubnikow:1930b,lifshitz1958theory} 
and the de Haas-van Alphen oscillations of the magnetic susceptibility~\cite{deHaas1930} 
have become wide-spread tools to characterize the Fermi liquid-like behavior of electrons in crystals. 

In contrast, the more challenging measurements of quantum oscillations in thermal conductivity 
are typically expected to bring little additional insight beyond Shubnikov-de Haas oscillations. 
The reason is that, in common metals, the electronic contribution to thermal transport is expected to satisfy the Wiedemann-Franz~(WF) law, which relates DC thermal conductivity $\kappa$  to the electrical one $\sigma$ as 
$\kappa{}=L_{0}T\sigma$, where $L_0=\frac{\pi^3}{3}(k_B/e)^2$ is the Lorentz number and $T$ the temperature. 
The proportionality between $\sigma$  and $\kappa$ reflects the fact that the same electronic degrees of freedom that carry charge  also carry heat, resulting in similar quantum oscillations. 

The situation is more involved in semimetals, for which the density of charge carriers can be several orders of magnitude smaller than in common metals such as copper. 
Whether the thermal transport is dominated by electrons, such as in metals, or by phonons, such as in insulators, is a subtle question. 
On one hand, the thermal conductivity of bismuth is dominated by the phononic contribution with a characteristic $T^{3}$ behavior at low temperatures~\cite{shalyt1944thermal,pratt1978thermal,Berman1976}. 
On the other hand, recent magneto-thermal transport experiments on a range of Weyl semimetals have displayed large enhancements of the Lorenz ratio interpreted as a non-trivial electronic contribution. 

In the Weyl semimetals NbP~\cite{Tanwar2022} and TaAs~\cite{Xiang2019} measurements of the longitudinal magneto-thermal conductivity have revealed the presence of quantum oscillations more than two orders of magnitude larger than expected based on the WF law. 
Due to the weak field dependence of phonon attenuation and presence of well defined chiral charge carriers it was suggested that the enhancement of magneto-thermal oscillations could originate from a novel collective excitation of the Fermi sea of chiral fermions, known as chiral zero sound~(CZS).
Interestingly, experimental work focusing on the transverse magnetothermal conductivity have also detected giant quantum oscillations. Here the apparent WF violation was interpreted as originating from ambipolar conduction accounting  for contributions of both electron- and hole-like excitations~\cite{stockert2017thermopower}. 
These experiments seem to discourage a general mechanism that leads to large thermal quantum oscillations applicable to all low-density semimetals. 

In this work, we put forward experimental and theoretical evidence of 
a phonon-electron scattering mechanism that leads to large magneto-thermal quantum-oscillations in all low-density semimetals, even when phonons dominate heat transport.
It rests on the peculiar phase-space constraints on the allowed phonon absorption mechanisms by electrons that occur when only a few Landau levels are filled.
As a consequence, magneto-thermal quantum oscillations have similar amplitudes and phase in directions along and perpendicular to the magnetic field. Besides the quantum oscillations, the varying phonon-absorption mechanisms parallel and perpendicular to the magnetic field $B$ lead to different $B$ dependences: a linear-in-$B$ increase of the thermal conductivity along the field, as opposed to a constant field dependence in directions perpendicular to the field.

We experimentally support this mechanism through a study of thermal transport in the Dirac semimetal \ch{ZrTe5}. 
Owing to its simple band structure, a single low-density band crossing the Fermi energy, we are able to quantitatively rationalize its giant quantum oscillations in thermal resistance, observed both in the longitudinal and transverse directions with respect to the magnetic field. 
Notably, the heat transport is phonon dominated, as we find a clear $T^{3}$ temperature dependence of the amplitude of quantum oscillations.
Because our analysis is based on phase-space arguments, we argue that it is generic to low-density metals, including bismuth and graphene.
More broadly, this mechanism seems indispensable to disentangle band-structure effects from strongly correlated phenomena in magneto-thermal quantum oscillations.
\begin{figure*}[th]
    \centerline{
    \includegraphics[width=0.9\textwidth]{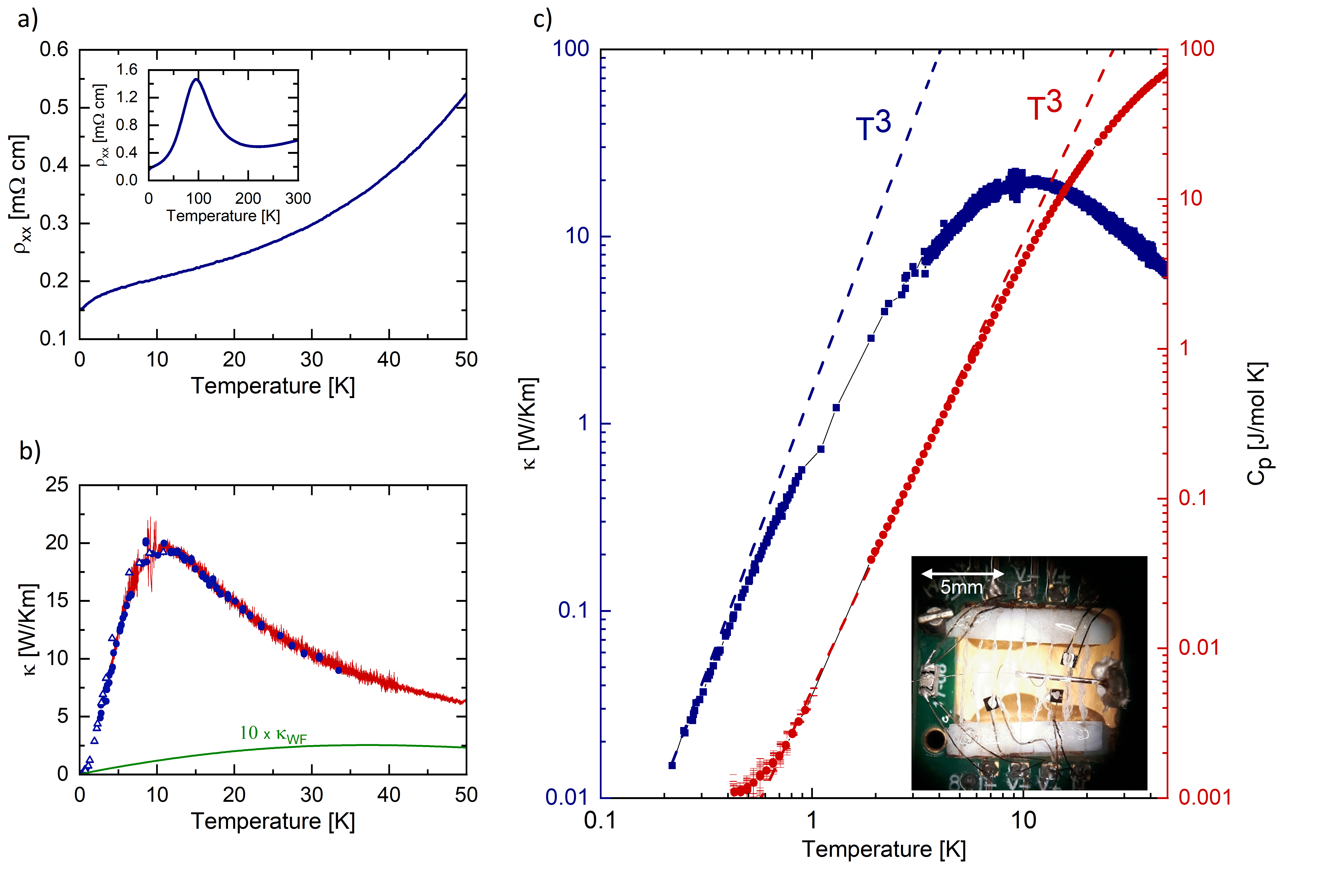}}
    \caption{{\bf Thermal and electrical transport in \ch{ZrTe5}.} (a) Electrical resistivity of \ch{ZrTe5} measured  with a $10$~$\mu$A current passed along the crystallographic $a$-axis. (b) Thermal conductivity of \ch{ZrTe5} measured with a thermal gradient ($\Delta{}T<0.1T$) applied along the $a$-axis. The continuous red line represents measurements where temperature was slowly changed and the gradient continuously recorded. Full circles and triangles represent measurements performed in two distinct cooldowns with temperature being stabilized for several minutes before taking the measurements. The solid green line represents an estimate of the thermal conductivity due to the electrons obtained from the Wiedemann-Franz law. (c) Comparison of thermal conductivity (blue points) and specific heat data (red points) and the expected $T^{3}$ dependence (dashed lines). Black hairlines act as a guide to the eye for the data. The inset displays a microscope photograph of one of the custom-build thermal conductivity setups. Cernox thermometers are attached to the sample using a $100$~$\mu$m silver wire. In this setup, used below 1K, superconducting TiN wires were used as electrical leads to the thermometers and heater. For measurements at higher temperatures long sections of manganin wire were used as electrical leads for thermal insulation.   
    }
    \label{fig1}
\end{figure*}
%
%-----------------------------------------------------%
\section{Magneto-thermal conductivity measurements}

In this study, we have selected \ch{ZrTe5} samples whose electronic 
properties and Fermi-surface topology have been thoroughly studied (for sample details see appendix of~\cite{galeski2021origin}).
One challenge in the study of transport properties of topological Weyl and Dirac semimetals lies in their often complex band structure~\cite{hasan2017}. The presence of different charge carrier types and multiple Fermi surface sheets of non-trivial shape often makes comparison between theoretical predictions and transport experiments demanding. 
One exceptional family of compounds are the pentatellurides \ch{ZrTe5} and \ch{HfTe5}. 
Multiple studies have shown that samples with low charge-carrier densities $n<10^{18}$~cm\textsuperscript{-3}, harbour only a single electron-like elliptical Fermi surface at the $\Gamma$ point comprising only few percent of the Brillouin zone and electron mobilities of the order of $400,000$~cm\textsuperscript{2}s\textsuperscript{-1}V\textsuperscript{-1}~\cite{tang2019three,galeski2022Lifshitz,Zhang2017,Zhang2021,Peng2020,Wang2020}. 
This makes the pentatellurides one of the simplest Dirac materials with only a single valley of electrons. 
Indeed, our recent study of electrical and thermoelectric transport, magnetization and sound propagation measurements have shown good agreement with linear response calculations based on the simple anisotropic Dirac Hamiltonian~\cite{galeski2021origin}. 
All measured samples harbour a small charge-carrier density of  approx. $1 \cdot 10^{17}$~cm$^{-3}$ and a single electron-like Fermi surface located at the $\Gamma$ point comprising less than $5\%$ of the Brillouin zone~(BZ). 
Thermal-transport data presented in this study has been measured using a home-made thermal-transport setup, see inset of Fig.~\ref{fig1}(c) (for details see Supplementary Material). 

As the first step towards understanding the thermal properties of \ch{ZrTe5}, we have compared its electrical (Fig.~\ref{fig1}(a)) with its thermal (Fig.~\ref{fig1}(b)) transport properties, when both heat and electrical currents are applied along the crystallographic $a$-axis. 
The measured thermal conductivity $\kappa$ displays a broad maximum at {\it ca.} $10$~K reaching the value of $\sim{}20$~W/Km, in good agreement with previous reports~\cite{Hooda2022,Chang-woo2022,Tritt2001,Zawilski2000}. 
This maximum is  attributed to the saturation of the phonons' mean free path due to scattering from crystal surfaces and large crystal defects such as grain boundaries \cite{Berman1976}. 
Comparison of the thermal conductivity with an estimate of the electronic contribution to the thermal transport derived from the WF law reveals, as expected for a low charge-carrier density metal, 
that the electronic contribution to the overall thermal conductivity is negligible (green line in Fig.~\ref{fig1}(b)).  
This is additionally confirmed by the temperature dependence of $\kappa$ at low temperatures (Fig.~\ref{fig1}(c)) where it follows the canonical $T^3$ temperature dependence expected for a thermal conductivity dominated by phonons~\cite{Berman1976}. 
Interestingly, our high resolution measurement does not reveal any signatures of phonon hydrodynamics as suggested in~\cite{Chang-woo2022}. 
In particular the slope of $\kappa$ never exceeds the canonical $T^3$ temperature dependence. 

Similarly, the measured specific heat shown in 
Fig.~\ref{fig1}c exhibits a $T^3$ dependence down to 500 mK suggesting a 
dominant role of the phonon degrees of freedom and a saturated mean free path. This is 
in reasonable agreement with Ref.~\cite{Behera2020}, and Shaviv {\it et al.} reported a slightly lower exponent~\cite{SHAVIV1989103}. 
At temperatures below 600mK a small excess in specific heat is observed. 
However, due to the exceedingly small specific heat of the samples compared to the calorimeter cell, those data are have a significant measurement uncertainty and will be discarded in further analysis.

\begin{figure*}[th]
    \centering
    \includegraphics[width=\textwidth]{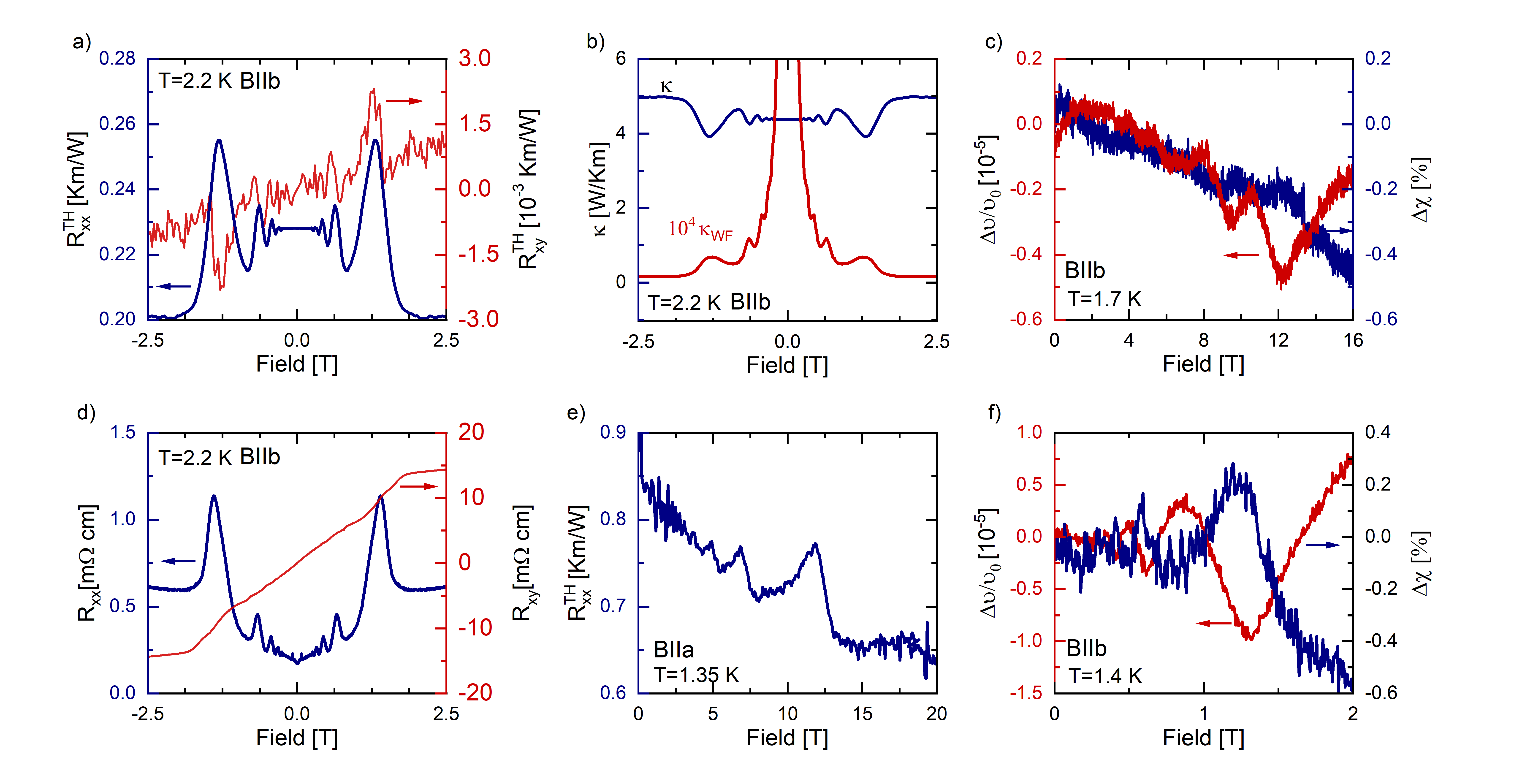}
    \caption{\textbf{Magnetotransport experiments on \ch{ZrTe5}.} (a) Symmetric and antisymmetric components of the thermal magneto-resistance tensor of \ch{ZrTe5} measured at 2.2K and with magnetic field applied parallel to the crystallographic \(b\)-axis and thermal gradient parallel to the \(a\)-axis. (b) Comparison of transverse thermal conductivity calculated from thermal magneto-resistance measurement (blue) and an estimate based on the Wiedemann–Franz law (red). (c) Sound–velocity variation \(\Delta v/v_0\) (red) and echo transmission amplitude (blue) of a longitudinal sound mode propagating along the \(a\)-axis as a function of magnetic field applied along the \(a\)-axis. (d) Symmetric and antisymmetric components of the magneto-resistance tensor of \ch{ZrTe5} measured at 2.2K and with magnetic field applied parallel to the crystallographic \(b\)-axis and electric current parallel to the \(a\)-axis. (e) Thermal magneto-resistance of \ch{ZrTe5} measured at 1.3K and with magnetic field and thermal gradient applied parallel to the crystallographic \(a\)-axis. (f) Sound–velocity variation \(\Delta v/v_0\) (red) and echo transmission amplitude (blue) of a longitudinal sound mode propagating along the \(a\)-axis as a function of magnetic field applied along the \(b\)-axis.}
    \label{fig2}
\end{figure*}

\begin{figure*}[th]
    \centerline{
\includegraphics[width=0.9\textwidth]{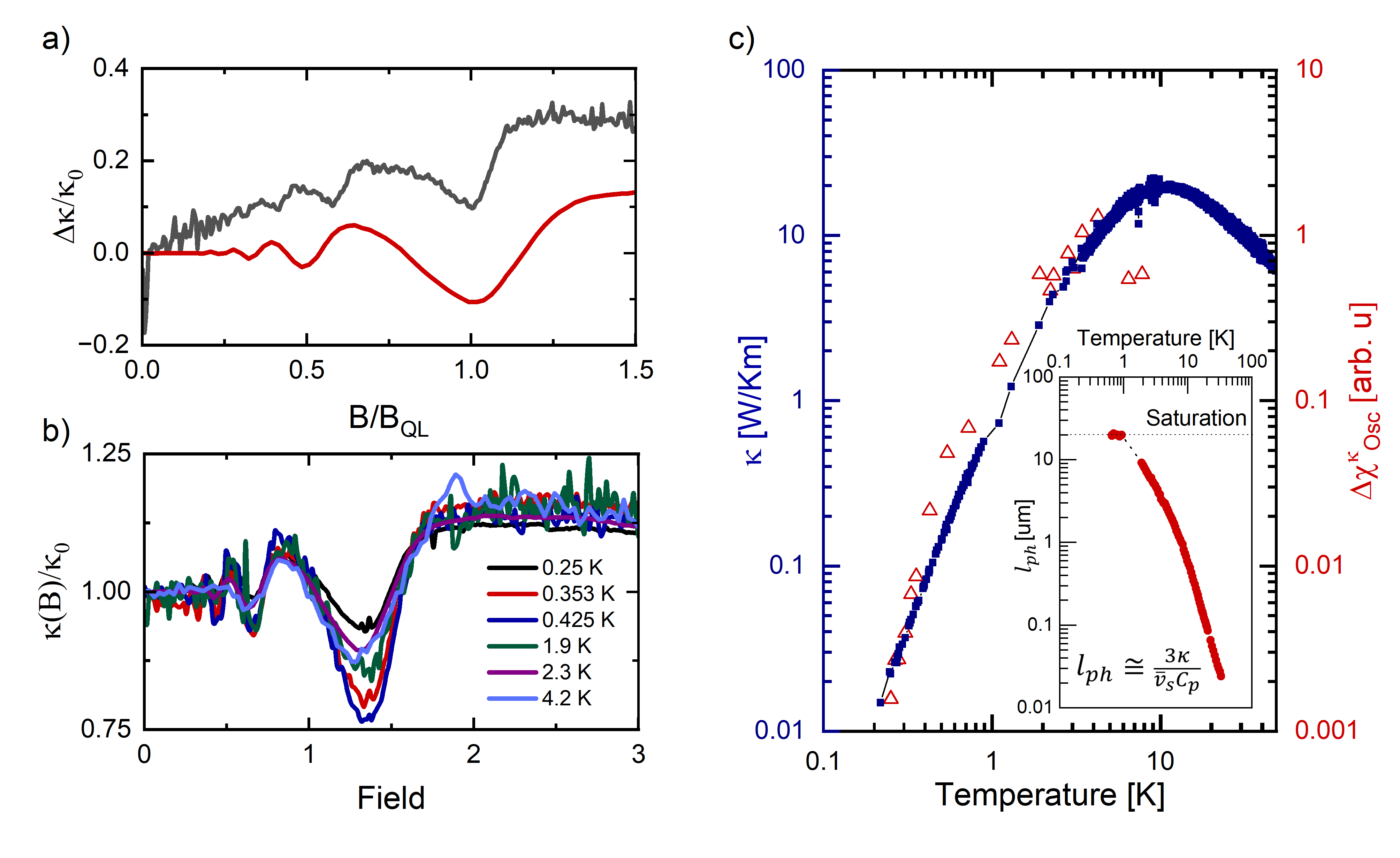}}
    \caption{{\bf Temperature dependence of the magneto-thermal quantum oscillation amplitude in \ch{ZrTe5}.}  (a) Comparison of transverse (red) and longitudinal (black) magneto-thermal 
    conductivity. For fields applied along the thermal gradient direction a clear linear contribution to
    conductivity is seen. The magnetic field strength is normalized to $B_{QL}$, the magnetic field strength above which only the lowest Landau level is occupied. (b) Temperature dependence of transverse magneto-thermal conductivity divided
    by $\kappa_0$. (c) Comparison of zero field thermal conductivity (blue) with the temperature
    dependence of the quantum oscillation amplitude. Both quantities follow a similar temperature
    dependence suggesting a common origin, inset shows the temperature dependence of the phonon mean free path. 
%    \noteAG{phonon mean free path in main text is called $l_{ph}$, so I would use the same label in the figure. $l_{m.f.p}$ is for electrons, later on. Update also equation in the inset to match \eqref{eq:kappaph}}. 
    }
    \label{fig3}
\end{figure*}

Although the thermal transport of \ch{ZrTe5} appears rather conventional  up to this point, 
further investigation of its field dependence unveils rather uncommon features. 
Fig.~\ref{fig2}(a) shows the magneto-thermal resistance and Righi-Leduc~(thermal Hall) effect with the magnetic field applied along the crystallographic $b$-axis measured at $2.2$~K. 
Inspection of the data reveals a negligibly small thermal Hall contribution to transport, on the edge of measurement sensitivity (red curve). 
In contrast, the longitudinal component of the thermal magneto-resistance exhibits giant oscillations matching the Shubnikov de-Haas oscillations observed in electrical resistance, Fig.~\ref{fig2}(d), measured on the same sample. 
The close resemblance of both effects suggests a common electronic origin of these oscillations, related to the Landau quantization of electronic orbits. 
This, however, is in stark contrast with the magnitude of the effect: the amplitude of the last oscillation amounts to almost $20\%$ of the total thermal resistance, even though the electronic contribution to thermal transport at zero field, estimated based on the WF law, is less than  $0.8\%$ of the total thermal conductivity. 
Indeed, comparison of the calculated thermal magneto-conductivity with an estimate based on the WF law, Fig.~\ref{fig2}(b), reveals that the amplitude of the measured thermal oscillations is about a factor $10^{4}$ larger than expected from an electronic contribution. 
In addition, the overall shape of the thermal magneto-conductivity differs significantly from the expected $B^{-1}$ behavior.

Moreover, quantum oscillations of the thermal conductivity of similar magnitude are also visible 
when both the thermal gradient and magnetic field are parallel to the a-axis, Fig.~\ref{fig2}(e). 
However, in this case, due to the presence of current jetting~\cite{Arnold2016,Pippard1989,Reis2016}, it is not possible to compare their magnitude with those of the electrical resistance (Appendix C for details). 
Interestingly, in this measurement configuration we observe, in addition to quantum oscillations, a negative contribution to the magneto-thermal resistivity (positive magneto-thermal conductivity, see discussion below). 

\begin{figure*}[!t]
    \centerline{
    \includegraphics[width=0.9\textwidth]{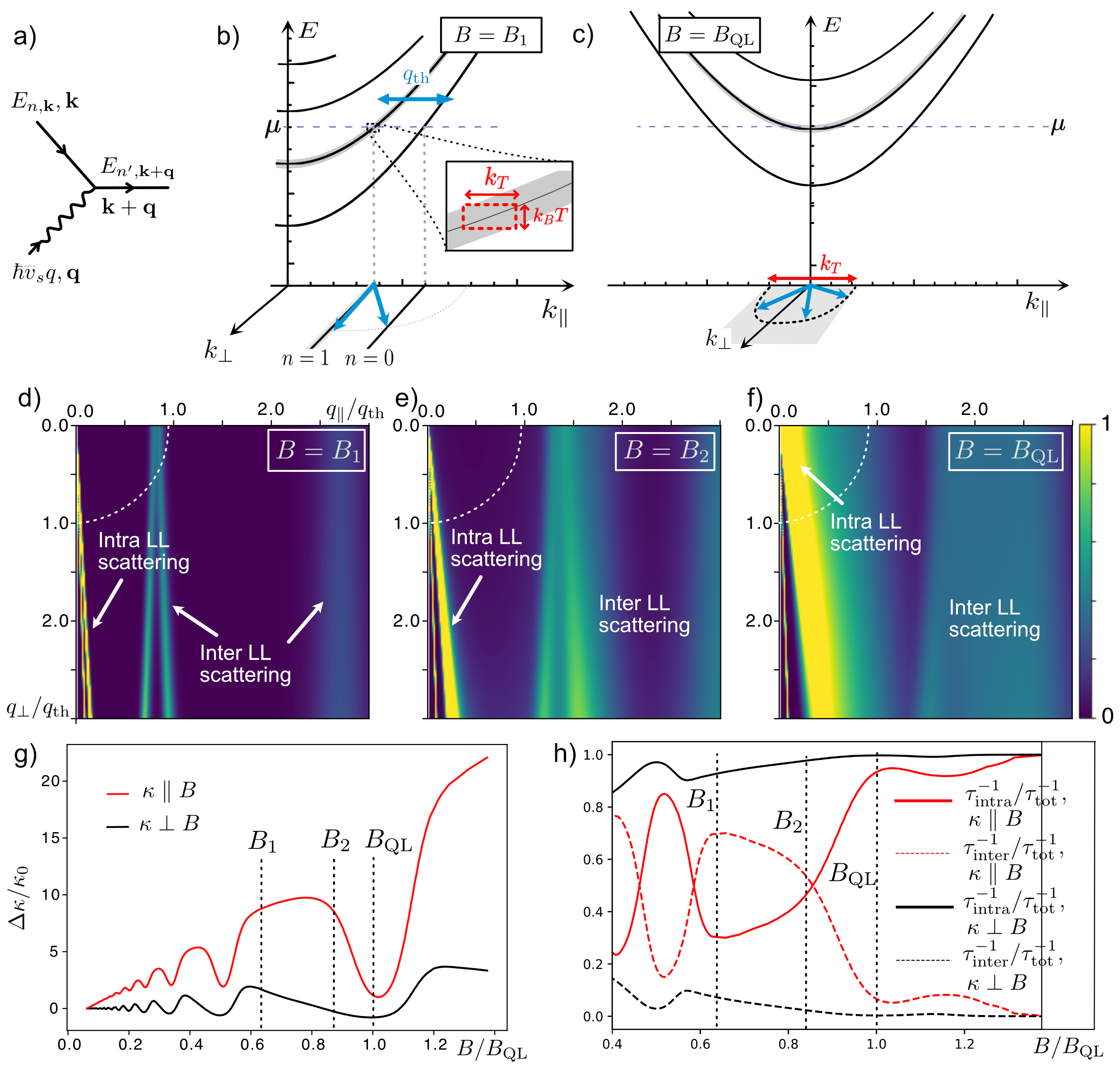}}
    \caption{%
    {\bf Quantum oscillations of thermal conductivity via phonon absorption by electrons.}
    \textcolor{black}{
    (a) Energy and momentum conservation constrain the absorption of a phonon of 
    momentum $\mathbf{q}$ by an electron between two LLs 
    $E_{n,\mathbf{k}}$ and $E_{n',\mathbf{k}+\mathbf{q}}$ 
%    as represented by the Feynman diagram in a). 
    (b) When the chemical potential lies away from the bottom of Landau levels, only a subset of phonon's momenta can be absorbed by electrons in inter-LL transitions and the phonon's absorption is 
    very anisotropic. 
 %   This leads to a very anisotropic phonon's absorption relative to the direction of the magnetic field. 
    The typical thermal phonon's momenta 
    $q_{\textrm{th}}=k_{\text{B}}T / \overline{v}_{\text{s}}$ for allowed processes are graphically represented. 
    (c) When the chemical potential crosses the bottom of a Landau level $n$, the absorption of phonons  
    in intra-LL transitions within the level $n$ becomes efficient and almost isotropic.   
    (d-f) The momentum dependence of the phonon's absorption rate $\hbar / \tau(\mathbf{q})$ 
%\noteAG{ add label in colorbar. I think it would be good to have each panel have its own letter, and refer to them sequentially in the main text.}
    is represented for different magnetic fields, for $T=2$~K, with the colors encoding its intensity. 
    (d,e) Away from the bottom of a Landau level, for $B=B_1,B_2$,  inter-LL processes dominate the phonon absorption in the direction of $\mathbf{B}$. 
    They absorb inefficiently phonons of momenta $q\gtrsim{}q_{\textrm{th}}$. 
    (f) At the bottom of a Landau level for $B=B_{\textrm{QL}}$, the absorption
    becomes dominated by efficient inter-LL transitions, manifested by a sudden increase of the absorption rate and a minimum of the thermal conductivity. 
%    In the direction perpendicular to $\mathbf{B}$, the absorption is always dominated by 
%    inter-LL transitions, much enhanced also at the bottom of a Landau level.
    (g) The resulting relative thermal conductivity 
    $\Delta{}\kappa(B) / \kappa_{0} = \tau(B) / \tau_{0}$ is plotted at $T=2$~K, 
    where  $\kappa_{0} = \kappa{}(B=0)$, in the direction parallel and perpendicular to $B$. 
    (h) 
    The inter and intra-LL transition's rates are plotted relative to the total absorption rate $\tau_{\textrm{tot}}^{-1} =\tau_{\textrm{inter}}^{-1} + \tau_{\textrm{intra}}^{-1}$ as a 
    function of the rescaled magnetic field $B/B_{\text{QL}}$, for a temperature $T=2$~K. 
    In the direction of $B$
    the inter-LL absorption mechanism dominates the absorption rate  away from the quantum oscillations, 
    giving rise to the $\kappa{}\propto{}B$ scaling in panel e).         
%    while at these oscillations the phonon's absorption is dominated by the  
%    intra-LL processes at the bottom of a Landau level. 
%    The inter-LL processes are at the origin of an average $\simeq{}B$ increase of the conductivity $\kappa{}\parallel{}B$ shown in panel e). 
    In contrast, perpendicular to the magnetic field, 
    intra-LL processes dominate the phonon's absorption irrespective of the 
    magnetic field, leading to an constant average  $\kappa$. 
    }
   }
    \label{fig4}
\end{figure*}

Although quantum oscillations are commonly attributed to  electrons and the existence of a Fermi surface, the dominance of the lattice degrees of freedom in thermal transport at zero field suggests that
the huge variations of the thermal conductivity in magnetic fields originate from changes of the 
phonon's mean free path due to the scattering on the electrons. 
To confirm the origin of the giant magneto-thermal quantum oscillations in \ch{ZrTe5}, we measured the 
temperature dependence of the magneto-thermal quantum oscillations down to $250$~mK.
Comparison of the temperature dependence of the amplitude of the last oscillation with the zero-field
thermal conductivity reveals that both quantities follow approximately a $T^{3}$ dependence, Fig.~\ref{fig3}(c), strongly suggesting that here, the quantum oscillations seen in thermal transport indeed are rooted in the phononic degrees of freedom. Indeed plotting $\kappa(B)$ divided by $\kappa(B=0)$ reveals good collapse of the data, Fig.~\ref{fig3}(b).

Recently, similar giant quantum oscillations have been reported in the longitudinal thermal conductivity
($\nabla T \parallel B$) of the Weyl semimetals TaAs and NbP~\cite{Xiang2019, Tanwar2022}. In these cases, the oscillations were attributed to the presence of CZS - a collective bosonic excitation of the Fermi sea of chiral fermions that requires multiple Fermi pockets. An alternative explanation was put forward for the presence of giant oscillations in  NbP in a 
transverse-field configuration ($\nabla{}T{}\perp B$):  the presence of both
electrons and holes was suggested to reduce the electrical conductivity without affecting the thermal conductivity, giving rise to a violation of the Wiedemann–Franz. 
Although both explanations seem feasible in the case of Weyl fermion materials harboring both electron and hole-like Fermi pockets, this is not the case in \ch{ZrTe5}. 
The samples studied in the present work have been shown~\cite{galeski2021origin,galeski2022Lifshitz} to contain 
only a single electron-like Fermi surface with a a massive Dirac dispersion. In addition with the Fermi level
of $\sim{}40-100$~K and the Dirac mass-gap of the order of $100$~K, an influence of thermally excited holes on
thermal transport  is excluded at the relevant temperatures, precluding the influence of both effects on thermal transport in the studied samples.

\section{Origin of the quantum oscillations in thermal transport.}
We now argue that a thermal current carried dominantly by phonons can display quantum oscillations originating from giant oscillations of the absorption rate of phonons by the electrons. 
%the variations of the electronic density of states. 
Moreover, we show that such oscillations appear both when the thermal current is directed parallel but also perpendicular to the magnetic field $\mathbf{B}$. 
Since a detailed theory of the phonons' thermal conductivity would be material specific, we resort to a general argument based on a phase-space argument to unveil  
the origin of these quantum oscillations. We argue there is a strong enhancement of the absorption of the thermal phonons by the electrons for specific values of the magnetic field. 
The phonon contribution to the thermal conductivity can be expressed as~\cite{abrikosov2017fundamentals}
\begin{equation}
\label{eq:kappaph}
%    \kappa_{\text{\text{ph}}} = \frac{1}{3}C_p\overline{v}_{\text{s}}l_{\text{\text{ph}}},
 \kappa = \frac{1}{3}C_p\overline{v}_{\text{s}}l_{\text{\text{ph}}},
\end{equation}
where $\bar{v}_s$ is the averaged phonon's velocity, the specific heat scales as $C_p\propto{}T^{3}$ at low temperatures, while at $\mathbf{B}=0$, the phonon mean free path $l_{\text{\text{ph}}}\simeq{}20$~$\mu$m is temperature independent, see inset of Fig.~\ref{fig3}(c). 
As the magnetic field is increased,  successive minima of the 
conductivity $\kappa{}(B)$, Fig.~\ref{fig3}(a), manifest a large decrease of this mean free path, signaling an enhanced absorption rate of phonons. 
By expressing $l_{\text{\text{ph}}} = \overline{v}_{\text{s}}\tau_{{\text{ph}}}$, we can sum the different contributions to the phonon scattering rate $\tau^{-1}_{{\text{ph}}}$ by resorting to the 
Matthiessen's rule. 
At low temperatures, the phonon-phonon Umklapp scattering is greatly reduced~\cite{ziman2001electrons}. 
The conduction electrons themselves constitute the main mechanism for scattering phonons, besides 
the sample surface and grain boundary scattering which is magnetic field independent. 

In the presence of a strong magnetic field $\mathbf{B}$, the dynamics of the electrons transverse to $\mathbf{B}$ freezes, leading to one-dimensional dispersive LLs
$E_{n,k_{\parallel}}$,  represented in Fig.~\ref{fig4}(b), indexed by $n$
notand the electron's momentum parallel to the field $k_{\parallel}$, here chosen along the $b$ crystallographic axis: 
%\noteAG{define $m$ and $v_{a,b,c}$}:
\begin{subequations}
\label{spectrum_ortho}
%\begin{align}
%    E_{0,k_{\parallel}} &= \pm{}\sqrt{(v_{b}\hbar{}k_{\parallel})^{2} +\left(m+\frac{1}{2}g\mu_{B}|B|\right)^{2}}\, , \\
%    E_{n,k_{\parallel}} &= \pm{}\sqrt{v_{b}^{2}\hbar^{2}k_{\parallel}^{2} +\left(\sqrt{m^{2}+2ne|B|\hbar{}v_{a}v_{c}}\pm{}\frac{1}{2}g\mu_{B}B\right)^{2}}
%    \ .
%\end{align}
\begin{align}
    E_{0,k_{\parallel}} &= \pm{}\sqrt{(v_{b}\hbar{}k_{\parallel})^{2}+m^{2}}\, , \\
    E_{n,k_{\parallel}} &= \pm{}\sqrt{(v_{b}\hbar{}k_{\parallel})^{2} +m^{2}+2ne|B|\hbar{}v_{a}v_{c}}
    \ .
\end{align}
\end{subequations}
with $v_{a,b,c}$ the Fermi velocity along the $a,b,c$ crystallographic axis  and $m$ 
a small mass parameter $m=10$meV. 
We neglected a subdominant Zeeman energy for the sake of clarity. 
Each of these levels has a density of states 
$\rho(B)= eB/(2\pi\hbar)$ associated to the different momenta $\mathbf{k}_{\perp}$. 
Thermal phonons have a typical energy $\hbar{}\omega{}\lesssim{}k_{\text{B}}T$ and momentum $|\mathbf{q}| \lesssim{}q_{\text{th}} =
k_B T / (\hbar \overline{v}_{\text{s}}) $.  
The absorption and emission of these phonons by the electrons is constrained by energy and momentum conservation which depend on the Landau levels, see Fig.~\ref{fig4}(a): 
\begin{equation}
    \mathbf{k}^{(\text{in})} +  \mathbf{q} =
    \mathbf{k}^{(\text{out})} 
    \ ; \ 
    E_{n,k^{(\text{in})}_{\parallel}} + \hbar{}\omega = E_{n',k^{(\text{out})}_{\parallel}} \ .
\end{equation}
In Fig.~\ref{fig4}(b) and (c), we represent schematically these conditions. 
%for the thermal phonon momentum $\mathbf{q}_{th}$ to be absorbed by an electron 
%with energy $E_n(\mathbf{k})$ into 
%$E_{n'}(\mathbf{k}+\mathbf{q}_{th})$. 
%
From Fermi's golden rule, the typical phonon's scattering scattering rate $\tau^{-1}_{\text{ph-e}}$ can be expressed as 
\begin{equation}
    \tau^{-1}_{\text{ph-e}} = 
    \int_{\mathbf{q}} ~  \tau^{-1} (\mathbf{q})
    / (\exp (\beta{}\hbar{}\overline{v}_{\text{s}}q )-1 ) , 
\end{equation}
with $\beta^{-1} = k_{\text{B}}T$ and  a scattering rate for phonons 
of momentum $\mathbf{q}$ that follows the energy and momentum conservation rules: 
\begin{multline}
    \hbar{}\tau^{-1}(\mathbf{q}) = 
    \sum_{n,n'}
    \int_{\mathbf{k}}
    f(E_{n,\mathbf{k}})
    (1-f(E_{n',\mathbf{k}+\mathbf{q}}))
    \\
    \delta{}( E_{n', \mathbf{k} +\mathbf{q}} - E_{n, \mathbf{k}} - \hbar\overline{v}_{\text{s}} |\mathbf{q}| )
    \left| 
    V_{\mathbf{k}+\mathbf{q};\mathbf{k},\mathbf{q}}
     \right|^{2},
\label{eq:ScattRate}
\end{multline}
%\begin{multline}
%    \hbar \tau^{-1}_\text{ph-e} (\mathbf{q}) = 
%    \int_{\mathbf{k}^{(\text{in})}}
%    \left< 
%%    f(E_{n,\mathbf{k}^{(\text{in})} })
%    (1-f(E_{n'} ( \mathbf{k}^{(\text{in})}+\mathbf{q})))
%    \right. 
%    \\
%    \left. 
%    \delta ( E_{n'} ( \mathbf{k}^{(\text{in})} + \mathbf{q}) - E_n ( %\mathbf{k}^{(\text{in})}) - \hbar \overline{v}_{\text{s}} |\mathbf{q}| )
%    \right>_{\mathbf{q}}
%    \left< \mathbf{k}^{(\text{out})} | V | \mathbf{k}^{(\text{in})} ; \mathbf{q})) %\right> |^2
%\end{multline}
with $f(E)$ the Fermi-Dirac distribution function. 
In a phase-space argument, we evaluate the variation of the scattering rate 
\eqref{eq:ScattRate} 
by neglecting the variation of the  matrix element of the electron-phonon coupling potential $V_{\mathbf{k}+\mathbf{q};\mathbf{k},\mathbf{q}}$, 
focusing on the energy and momentum constraints. 
The dependence on the phonon momentum $\mathbf{q}$ of this scattering rate is represented in Fig.~\ref{fig4}(d-f) for different values of the magnetic field 
$B$, see App.~\ref{app:Numerical} for details on the numerical method. 

%\noteAG{This paragraph is quite important but I find it quite confusing. I tried iterating it a little but should be revisited by David. Perhaps it is good to label the different panels in 4.d with their own letter and refer to them sequentially.  In fact, I find the caption is much clearer. }

The strong anisotropy of the Landau levels is reflected in how different  mechanisms dominate the absorption in various directions of phonon propagation. 
Along the magnetic field, only 
a few discrete momenta $q_{\parallel}$ can be absorbed via \emph{inter-LL} transitions 
away from the bottom of a LL, see Fig.~\ref{fig4}(b) and (d). 
This results in a rather inefficient absorption at high magnetic fields and a peak in the thermal conductivity, see Fig.~\ref{fig4}(g). 
The absorption rate along the field is thus set by the number of relevant inter-LL transitions. 
The number of LLs~\eqref{spectrum_ortho} crossing the chemical potential scales as $\propto{}1/B$, leading to a number of  inter-LL transitions scaling as $1/B^{2}$. Hence, accounting for the electronic density of LLs, the resulting average absorption rate \eqref{eq:ScattRate} scale as $\hbar \bar{\tau}^{-1}\propto 1/B$, as shown on the net linear 
increase in field of $\kappa \parallel B$ shown on Fig.~\ref{fig4}(g)
In the direction perpendicular to $\mathbf{B}$, the absorption is dominated by 
\emph{intra-LL} transitions, see Fig.~\ref{fig4}(d,e), and it is very directional, as shown in Fig.~\ref{fig4}(h). 
As a consequence,  the absorption rate perpendicular to the field is almost constant for fields away from the bottom of LLs, leading 
to an average constant  $\kappa \perp B$ as opposed to the linear 
increase of $\kappa \parallel B$ shown on Fig.~\ref{fig4}(g). 

At the bottom of each Landau band, where $\mu{}\simeq{}E_{n,\mathbf{k}=0}$, the absorption mechanism becomes dominated by intra-LL transitions within the  level $n$ both along and perpendicular to $\mathbf{B}$. Simultaneously, absorption of small phonon momenta in all directions become possible, leading to a sudden increase of the efficiency of this absorption, as shown in Fig.~\ref{fig4}(f). 
This sudden increase of the absorption rate $\hbar{} / \tau_{\textrm{tot}}$ for each LL leads to the observed 
\emph{quantum oscillations of the phonon-dominated thermal conductivity}, 
as shown in the theoretical curves of Fig.~\ref{fig4}(g).

\section{Discussion}

We have shown that thermal conductivity dominated by phonons can display giant quantum oscillations due to the increased phonon absorption by electrons at the bottom of Landau levels. One could agrue that presence of such an effect could be corroborated by direct measurements of the sound attenuation. Indeed,
while its relation to thermal conductivity was not discussed, oscillations in the sound absorption in relation to Landau levels for magnetic fields parallel to the sound propagation direction were discussed as early as the 1960s~\cite{gurevich1961giant,shapira1965line}, see also Sec. 12.8 of~\cite{abrikosov2017fundamentals}.

To contrast the experimental manifestation of quantum oscillations in both quantities, 
we have conducted sound attenuation measurements in our samples, Fig.~\ref{fig2}(c) and \ref{fig2}(f). 
They reveal that the relative changes in the attenuation and the speed of sound due to magneto-acoustic quantum oscillations do not exceed $1\%$, 
This appears at odds with the proposed attenuation mechanism. Indeed, recently it has been argued~\cite{Xiang2019} that since attenuation measurements probe only the phonon degrees of freedom, it is expected that strong quantum oscillations appearing in thermal conductivity due to phonon attenuation by electrons should manifest as a strong variation of the echo amplitude. 

This apparent contradiction is resolved by 
%considering the difference of wavelengths of the propagating phonons in both situations, since 
realizing that thermal-conductivity and sound-attenuation measurements 
%in typical solid state experiments 
probe very different energy scales. 
In  ultrasonic measurements one typically probes the phonons with frequencies $\lesssim{}1$~GHz, whereas
thermal conductivity probes thermal phonons whose frequencies, even at $200$~mK, exceed  $25$~GHz.
Attenuation of phonons by electrons can be separated into two regimes: the low-frequency "hydrodynamic"
regime where the phonon wavelength is much larger than the electron mean free $q\cdot{}l_{m.f.p}\ll{}1$ path and attenuation increases quadratically with phonon frequency, the other regime is the "quantum limit" of attenuation $q\cdot{}l_{m.f.p}\gg{}1$ where attenuation increases linearly with frequency~\cite{luthi2007physical}. 
In the case of \ch{ZrTe5}, the sound wavelength at $314$~MHz is $\sim{}10$~$\mu$m, and the electronic mean free path at $\lesssim{}2$~K is of the order of $1$~$\mu$m placing the ultrasound measurement in the hydrodynamic limit. In contrast typical phonons probed in thermal conductivity at $200$~mK have a wavelength of $\sim{}0.1$~$\mu$m, well in the quantum limit of absorption. This makes phonon absorption at the frequencies probed by thermal conductivity measurements more than two orders of magnitude stronger than in typical ultrasound experiments, making it a much more sensitive technique to study electron-phonon processes~\cite{Holland968}. Thermal conductivity appears as an ideal probe of the giant quantum oscillations of the phonon absorption rate by electrons.

Although, for clarity, we have focused our experiments and modeling on one of the simplest Dirac semimetals, \ch{ZrTe5}, there is ample experimental evidence suggesting the generality of the proposed mechanism in semimetals whose thermal transport is dominated by phonons. 
Signatures of quantum oscillations in $\kappa$, tentatively attributed to phonon absorption, have been seen in Bi~\cite{steele1955oscillatory} at magnetic fields where only a few LL were occupied. 
Similarly, quantum fluctuations of the phonon-dominated thermal conductivity of Sb were attributed to fluctuations in the number of scattering centers for phonons in~\cite{long1965electron}. 
In addition, here the longitudinal thermal conductivity displayed a linear in $B$ behavior besides quantum fluctuations, characteristic of the inter-LL absorption processes described in the present paper. 
More recently, studies on Sb~\cite{Jaoui2022} have also found a regime where the thermal conductivity was dominated by phonons, yet displayed quantum oscillations of their thermal conductivity; this was attributed to a momentum exchange between phonons and electrons. In addition, weak quantum oscillations and a linear-in-$B$ dependence were also observed in the thermal conductivity of graphite~\cite{ayache1980quantum} and graphene~\cite{Crossno2017}. 

Recently thermal-conductivity measurements have taken a central stage in the study of quantum matter due to their potential to reveal exotic low-energy excitation's such as charge neutral modes (see e.g.~\cite{Wakeham2016,Sato2016,Kasahara2019,sato2019}).

Traditionally in thermal transport measurements one seeks additional non-trivial contributions to thermal conductivity on top of a simple phononic background. Surprisingly, recent studies of the thermal Hall effect have shown that scattering of phonons from electronic or spin subsystem can lead emergence of non-trivial contributions in the phononic thermal Hall conductivity. ~\cite{balents2022hall,behnia2020Hall,Behnia2023Hall,Sachdev2022Hall,Taillefer2022Hall}. Here we show that the more traditional longitudinal phononic thermal transport can also acquire a non-trivial field dependence due to scattering from electronic excitations. The proposed phase-space mechanism is general and should be applicable to any thermal measurement in magnetic fields, including the measurements needed to interpret more exotic phenomena like the Thermal Hall.

%
% This quantity is much more accessible than the thermal Hall effect. 
%
%In fact observation of pronounced
%quantum oscillations from an exceedingly dilute Fermi sea 
%suggests that measurements of the field dependence of diagonal components of phononic thermal conductivity could be an
%efficient probe for detection of other exotic quasi-particles. % that could be otherwise 

\section{Acknowledgements}

We thank K. Behnia, B. Gotsmann, B. Fauqué, S. Hartnoll and J. Kroha for inspiring conversations that started this work. A.G.G. acknowledges financial support from the European Union Horizon 2020 research and innovation program under grant agreement No. 829044 (SCHINES). The work at BNL was supported by the US Department of Energy, oﬃce of Basic Energy Sciences, contract no. DOE-sc0012704. This work was supported by HLD-HZDR, member of the European Magnetic Field Laboratory (EMFL).

\appendix
\section{Materials and methods}

\subsubsection{Sample synthesis and preparation}
High-quality single-crystalline \ch{ZrTe5} samples were synthesized using high purity elements ($99.9999\%$ zirconium and $99.9999\%$ tellurium). Needle-shaped crystals (about  $0.2\times{}0.3\times{}3$~mm$^{3}$) were obtained by a tellurium flux method. Prior to transport measurements, Pt contacts were sputter deposited on the sample surface to ensure low contact resistance. The contact geometry was defined using Al hard masks. This procedure allowed us to achieve contact resistances below  $1$~$\Omega$. Band-structure parameters of the main sample have been studied in Ref.~\cite{galeski2021origin}, where it is refereed to as sample B. 

\subsubsection{Sample enviroment}
All transport measurements up to $9$~T were performed in a temperature-variable cryostat (PPMS Dynacool, Quantum Design), equipped with a dilution refrigerator insert. Additional longitudinal thermal transport experiments were performed in a 22 T Oxford instruments cryostat equipped with a $^{3}$He insert. In addition ultrasound  measurements were performed in a $16$~T Oxford instruments cryostat equipped with a standard Variable Temperature Insert (VTI) and Kelvinox MX400 Dilution refrigeratos.

\subsubsection{Ultrasound experiments}
Ultrasound measurements were performed using a phase-sensitive pulse-echo technique. Two piezoelectric lithium-niobate (LiNbO$_{3}$) resonance transducers were glued to opposite parallel surfaces of the sample to excite and detect acoustic waves. The sample surfaces were polished using a focused Ion beam in order to ensure that the transducer attachment surfaces were smooth and parallel. The longitudinal acoustic waves were propagated along the $a$-axis. Relative sound-velocity  $\Delta{}v/v$, and sound attenuation $\Delta\alpha$, were measured for field applied along the $a$. Data showing propagation with field along the $b$-axis have been previously published in ref~\cite{galeski2021origin}.

\section{Numerical method}
\label{app:Numerical}

For the numerical evaluation of the phonon-scattering rates for phonons of momentum $\textbf{q}$, associated to a transfer of an electron from Landau level $n$ to Landau level $n'$ \eqref{eq:ScattRate}, we first perform the integration on the variable $k_{\perp}$, leading to a density of states per Landau level equal to $\frac{e|B|}{2\pi\hbar}$. Then, to integrate over $k_{\parallel}$, we determined the zeros of
\begin{equation}
    \label{eq:energies}
    f_\mathbf{q}(k_{\parallel})=E_{n',k_{\parallel} +q_{\parallel}} - E_{n,k_{\parallel}} - \hbar \overline{v}_{\text{s}} |\mathbf{q}|
\end{equation}
by using a dichotomy algorithm. The integral can then be evaluated by using the property of the $\delta(x)$ distribution, for a function $f(x)$ vanishing at positions $x\in\left\lbrace x_{0},x_{1},...,x_{n}\right\rbrace$ 
\begin{equation}
    \delta(f(x))=\sum_{k=0}^{n}\frac{\delta(x-x_{k})}{\left|\left.\partial_x f\right|_{x=x_{k}}\right|}\,.
\end{equation}
After summation on $n$ and $n'$, the result of these integrations at a $T=2K$ is plotted in Fig.~\ref{fig4}(d,e,f) for three different values of magnetic-field.\\

Afterwards, it is possible to determine the total scattering rate by evaluating the integral over the phonon momentum $\textbf{q}$, taking into account the thermal occupation of the phonon states captured by a Bose-Einstein function.\\

The experiment measured longitudinal and transverse transport. To capture the difference in between these two situations numerically, we defined $\tau_{\perp}$ and $\tau_{\parallel}$, such that $\kappa^{\parallel/\perp}=\frac{1}{3}C_p\Bar{v}_{\text{s}}^{2}\tau_{\parallel/\perp}$, and evaluated them by assuming that the phonons participating in the transverse and longitudinal transport measurement are those contained in a solid angle $\frac{\pi}{6}$ around the transport direction, such for example that 
\begin{equation}
    \tau^{-1}_\text{ph-e,$\parallel$} = 
    \int_\mathbf{q} ~    \frac{\tau^{-1}(\mathbf{q})}{\exp{}(\beta{}\hbar{}\overline{v}_{\text{s}}q ) -1}\theta(q_{\parallel}-2q_{\perp}) \,, 
\end{equation}
with $\theta(x)$ the Heaviside function.

\section{Additional Data}
\subsection{Field dependence of specific heat of \texorpdfstring{\ch{ZrTe5} }crystals.}

\begin{figure}[!ht]
    \centerline{
    \includegraphics[width=0.5\textwidth]{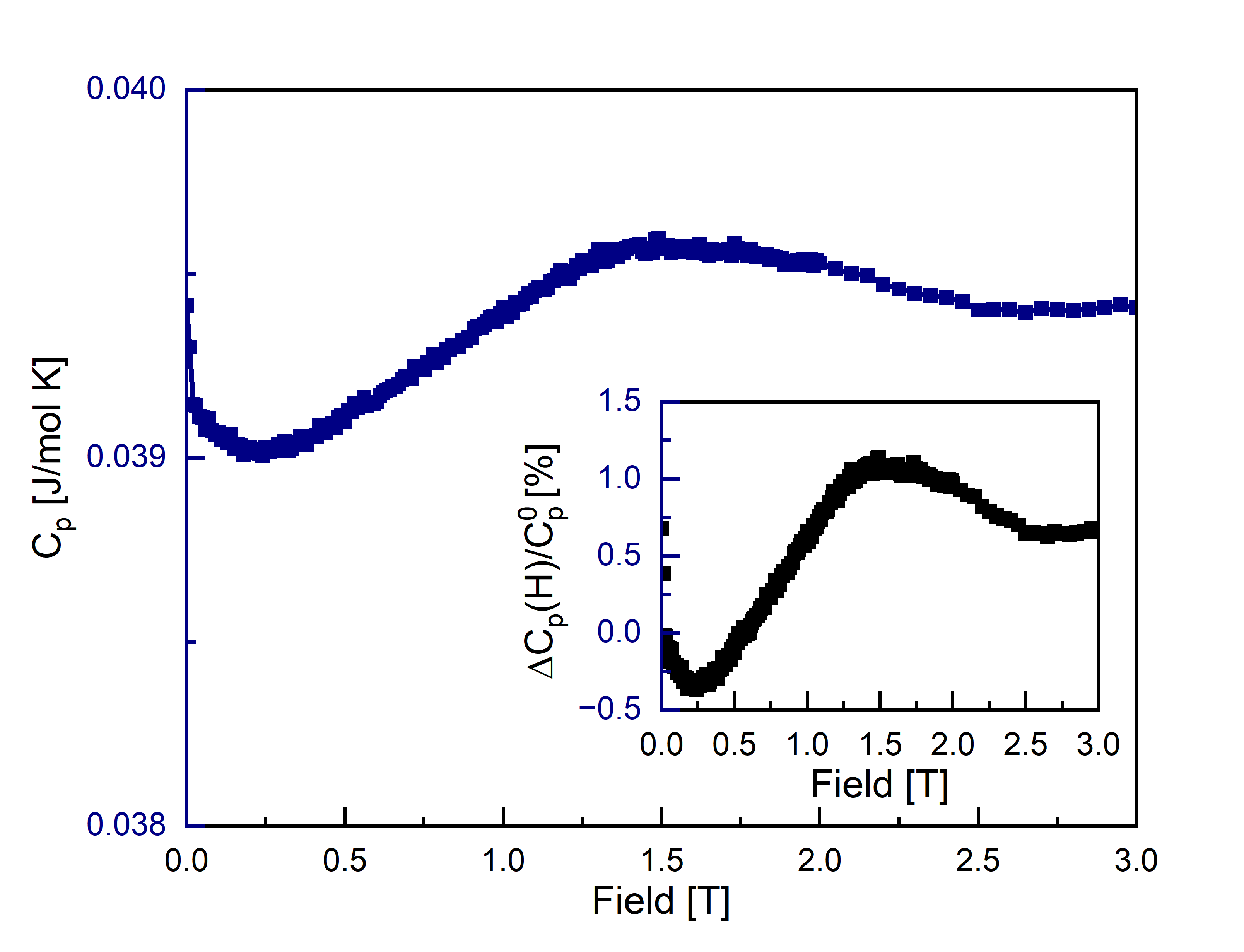}}
    \caption{Field ($B\parallel$ $b$-axis ) dependance of the specific heat measured at $1.9$~K. The inset shows the change of $C_{p}$ in percent. The field induced changes amount to maximum $1\%$ of the total specific heat and, thus, cannot account for the huge quantum oscillations seen in thermal transport.)  
    }
    \label{Cp}
\end{figure}

In our analysis, we have assumed that the specific heat of \ch{ZrTe5} is approximately independent of magnetic field and, thus, the variations of the  phonon thermal conductivity  $\kappa = \frac{1}{3}\cdot{}C_{\text{p}}{}\cdot{}\overline{v}_{\text{s}}\cdot{}l_{\text{ph}}$ in magnetic field are given by changes of $l_{\text{ph}}$. To verify if this assumption holds we have measured the specific heat of \ch{ZrTe5} at $2$~K using a standard PPMS specific heat option, using a field calibrated puck. The results shown in Fig.\ref{Cp} confirm our assumption. Field-induced variations of $C_{\text{p}}(H)$ do not exceed $1\%$ and, thus cannot account for the almost $20\%$ changes of $\kappa$. In particular, $C_{\text{p}}(H)$ does not exhibit any signs of quantum oscillations. 

\subsection{Signatures of current jetting in longitudinal charge transport measurements.}

\begin{figure}[!ht]
    \centerline{
    \includegraphics[width=0.6\textwidth]{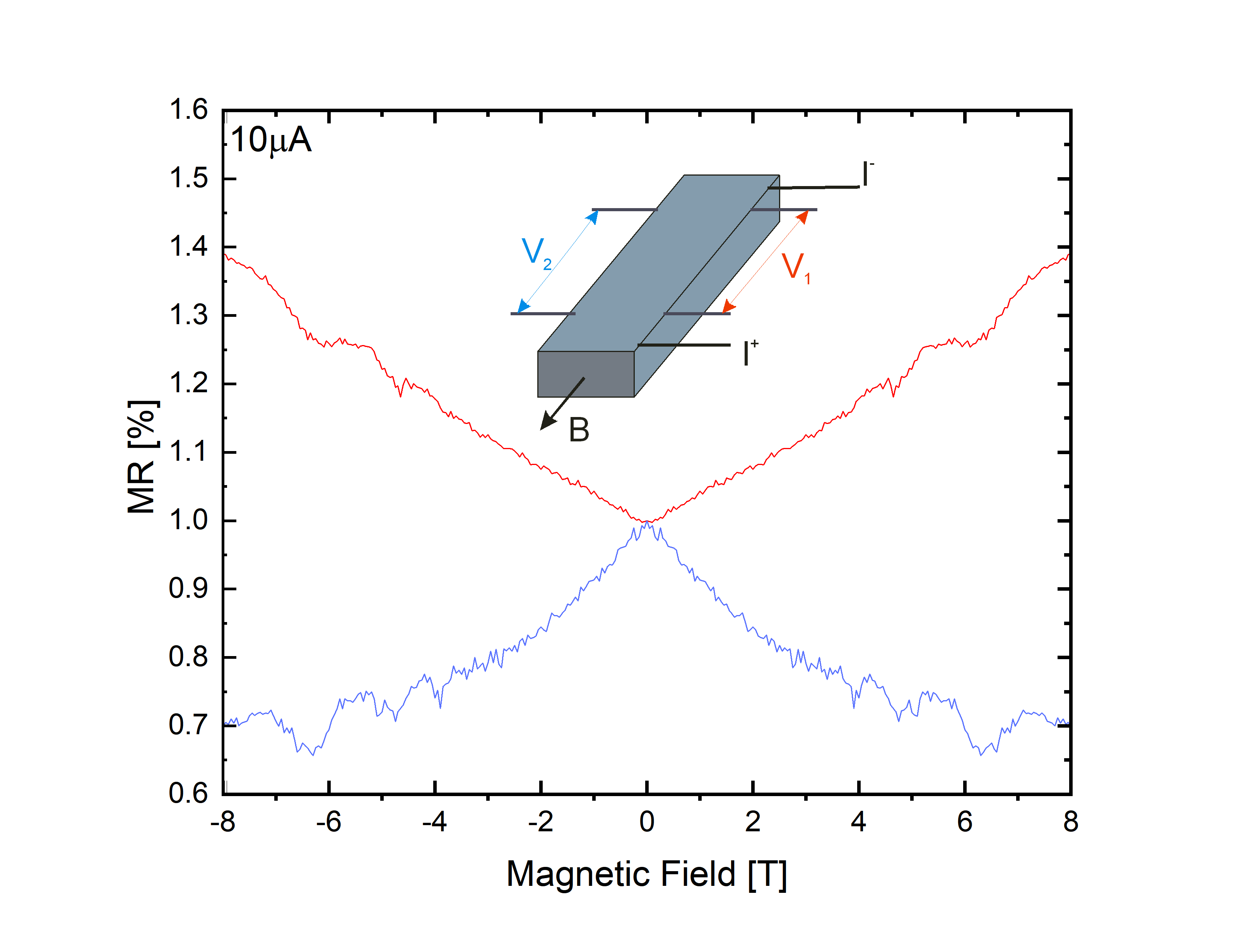}}
    \caption{Signatures of current jetting in longitudinal charge transport measurements in \ch{ZrTe5}. The magnetic field and electric current are both applied along the a-axis. In this experiment current was applied via contacts located on the left side of the crystal. Radically different field-dependant voltage drops mat the contacts V$_{1}$ and V$_{2}$ evidence that the current distribution in this measurement configuration is not homogeneous.)  
    }
    \label{jetting}
\end{figure}

Current jetting is a phenomenon, where the electrical current distribution within a sample is highly non-uniform in measurements of the longitudinal magneto-resistance. In particular, in samples with low charge carrier density 
current does not "even out" after injection through imperfect contacts and can form "jets" of current along the direction of the magnetic field. 
This  uneven distribution of current can lead to misleading results in experiments such as appearance of negative magnetoresistance~\cite{Pippard1989,Arnold2016,Reis2016}. One method for determining weather current jetting is present in resistance measurements is by applying current to a pair of contacts located on one side of the sample and comparing voltage drops measured on the same and opposite sides  of the sample~\cite{Arnold2016} as, see sketch of the measurement geometry in Fig. \ref{jetting}. Results of such a measurement performed on \ch{ZrTe5} samples at $2$~K are shown in Fig. \ref{jetting}. Measurements of the voltage drop on opposite sides of the sample display a dramatically different field dependence strongly suggesting presence of current jetting and making longitudinal magneto-resistance measurements unreliable. Current jetting is not relevant in measurement of the thermal conductivity since there are no current injection point in such measurement.

\bibliography{Biblio}% Produces the bibliography via BibTeX.

\end{document}

\bibliographystyle{apsrev4-1}
\title{
Supplementary information}

\author{B. Bermond}
\email{baptiste.bermond@ens-lyon.fr}
\affiliation{ENSL, CNRS, Laboratoire de Physique, F-69342 Lyon, France}

\author{R. Wawrzy\'{n}czak}
\affiliation{Max Planck Institute for Chemical Physics of Solids, Nöthnitzer Stra{\ss}e 40, 01187 Dresden, Germany}

\author{S. Zherlitsyn}
\affiliation{Hochfeld-Magnetlabor Dresden (HLD-EMFL) and Würzburg-Dresden Cluster of Excellence ct.qmat, Helmholtz-Zentrum Dresden-Rossendorf, 01328 Dresden, Germany}

\author{T. Kotte}
\affiliation{Hochfeld-Magnetlabor Dresden (HLD-EMFL) and Würzburg-Dresden Cluster of Excellence ct.qmat, Helmholtz-Zentrum Dresden-Rossendorf, 01328 Dresden, Germany}

\author{T. Helm}
\affiliation{Hochfeld-Magnetlabor Dresden (HLD-EMFL) and Würzburg-Dresden Cluster of Excellence ct.qmat, Helmholtz-Zentrum Dresden-Rossendorf, 01328 Dresden, Germany}

\author{D. Gorbunov}
\affiliation{Hochfeld-Magnetlabor Dresden (HLD-EMFL) and Würzburg-Dresden Cluster of Excellence ct.qmat, Helmholtz-Zentrum Dresden-Rossendorf, 01328 Dresden, Germany}

\author{G. D. Gu}
\affiliation{Condensed Matter Physics and Materials Science Department, Brookhaven National Laboratory, Upton, NY, USA}

\author{Q. Li}
\affiliation{Condensed Matter Physics and Materials Science Department, Brookhaven National Laboratory, Upton, NY, USA}

\author{F. Janasz}
\affiliation{Department of Engineering, Faculty of Science, Medicine and Technology, University of Luxembourg, Luxembourg}

\author{T. Meng}
\affiliation{Institute for Theoretical Physics and Würzburg-Dresden Cluster of Excellence ct.qmat, Technische Universität Dresden, Dresden, Germany}

\author{F. Menges}
\affiliation{Max Planck Institute for Chemical Physics of Solids, Nöthnitzer Stra{\ss}e 40, 01187 Dresden, Germany}

\author{C. Felser}
\affiliation{Max Planck Institute for Chemical Physics of Solids, Nöthnitzer Stra{\ss}e 40, 01187 Dresden, Germany}

\author{ J. Wosnitza}
\affiliation{Hochfeld-Magnetlabor Dresden (HLD-EMFL) and Würzburg-Dresden Cluster of Excellence ct.qmat, Helmholtz-Zentrum Dresden-Rossendorf, 01328 Dresden, Germany}
\affiliation{Institut für Festkörper-und Materialphysik and Würzburg-Dresden Cluster of Excellence ct.qmat, Technische Universität Dresden, 01062 Dresden, Germany}

\author{Adolfo G. Grushin}
\email{adolfo.grushin@neel.cnrs.fr}
\affiliation{Univ. Grenoble Alpes, CNRS, Grenoble INP, Institut N\'eel, 38000 Grenoble, France}

\author{David Carpentier}
\email{david.carpentier@ens-lyon.fr}
\affiliation{ENSL, CNRS, Laboratoire de Physique, F-69342 Lyon, France}

\author{ J. Gooth}
\affiliation{Max Planck Institute for Chemical Physics of Solids, Nöthnitzer Stra{\ss}e 40, 01187 Dresden, Germany}
\affiliation{Physikalisches Institut, Universität Bonn, Nussallee 12, 53115 Bonn, Germany}

\author{S. Galeski}
\email{sgaleski@uni-bonn.de}
\affiliation{Physikalisches Institut, Universität Bonn, Nussallee 12, 53115 Bonn, Germany}
\affiliation{Max Planck Institute for Chemical Physics of Solids, Nöthnitzer Stra{\ss}e 40, 01187 Dresden, Germany}
\affiliation{Hochfeld-Magnetlabor Dresden (HLD-EMFL) and Würzburg-Dresden Cluster of Excellence ct.qmat, Helmholtz-Zentrum Dresden-Rossendorf, 01328 Dresden, Germany}

%%%%%%%%%%%%%%%%%%%%%%%%%%%%%%%%%%%%%%%%%%%%%%%%%%%%%%%%%%%%%%%%%%%%%%%%%%%%%
%
%\date{\today}
% \date{June 17, 2022}
\begin{abstract}
\end{abstract}
\maketitle

\section{ Estimating the temperature distribution across the thermal transport setup }

In this work we have used a custom build thermal transport setup. In order to estimate heat flow direction and estimate potential heat losses to gauge whether the sample is appropriately isolated form the outside world we have simulated the temperature distribution across the setup using a finite elements thermal transport solver included in the ANSYS package. In the computation we have included radiation effects and used material parameters tabulated in Table 1 reflecting the materials used in design of the experiment.

\begin{table}
    \centering
    \begin{tabular}{lrrr}
         \textbf{Material}&  \textbf{Density}&  \textbf{Thermal conductivity}& \textbf{Heat capacity}\\
         &  \textit{g/cm3}&  \textit{W/mK}& \textit{J/kgK}\\
         Manganin&  8.4&  0.0335& 0.8\\
         Sapphire&  3.98&  2.4& 0.02\\
         Silicon&  2.3&  200& 0.005\\
         Silver&  10.49&  3940& 1\\
         Steel&  7.85&  60.5& 434\\
         ZrTe5&  5.6&  0.7& 0.005\\
    \end{tabular}
    \caption{Material properties used in ANSYS simulations}
    \label{matProps}
\end{table}

\begin{figure*}[th]
    \centerline{
    \includegraphics[width=1\textwidth]{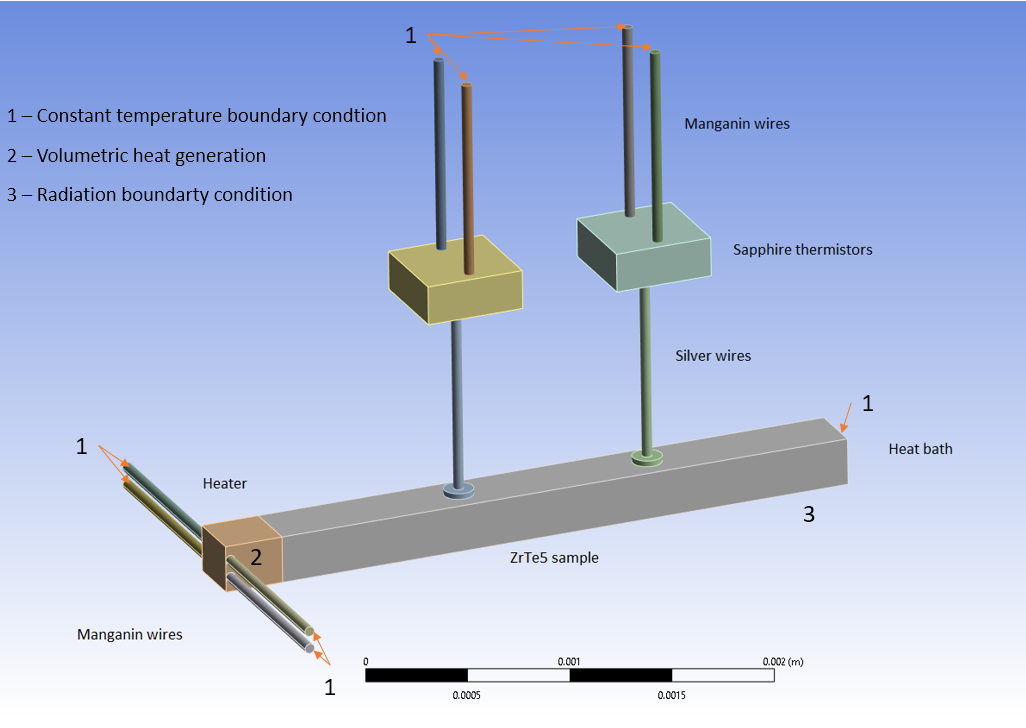}}
    \caption{Thermal boundary conditions applied   
    }
    \label{BC}
\end{figure*}

Thermal conductance of the manganin wires has been adjusted to reflect thermal conductance of the real length of used wire used in the experiment. 
The calculation domain shown in Fig.\ref{BC} is composed of the ZrTe5 sample, two thermistors and a heater, connected with silver and manganing wires respectively - reflecting the used experimental setup. Constant temperature boundary condition was applied to the end of the sample, representing a connection to the heat bath and to end of each wire. In addition, radiative boundary condition was applied to the ZrTe5 sample to simulate the heat losses. Finally, volumetric heat generation was included in the heater element, to simulate the Joule heating used in the experiment.  Table 2 shows a summary of used values.

\begin{table}
    \centering
    \begin{tabular}{|l|r|r|} \hline 
         \textbf{Heating BC}&  Heat generation rate& Equivalent power\\ \hline 
         &  \textit{W/m3}& \textit{uW}\\ \hline 
         Simulation 1&  0.116e4& 0.025\\ \hline 
         Simulation 2&  4.17e4& 0.9\\ \hline 
         Simulation 3&  11.6e4& 2.5\\ \hline 
         \textbf{Radiation BC}&  Emissivity& Ambient temperature\\ \hline 
         &  & \textit{K}\\ \hline 
         In all simulations&  0.4& 2\\ \hline 
         \textbf{Temperature BC}&  Temperature at contacts& \\ \hline 
         & \textit{K}& \\ \hline 
         In all simulations&  1& \\ \hline
    \end{tabular}
    \caption{Applied boundary conditions}
    \label{boundaryConditionsTable}
\end{table}

\begin{figure*}[th]
    \centerline{
    \includegraphics[width=0.6\textwidth]{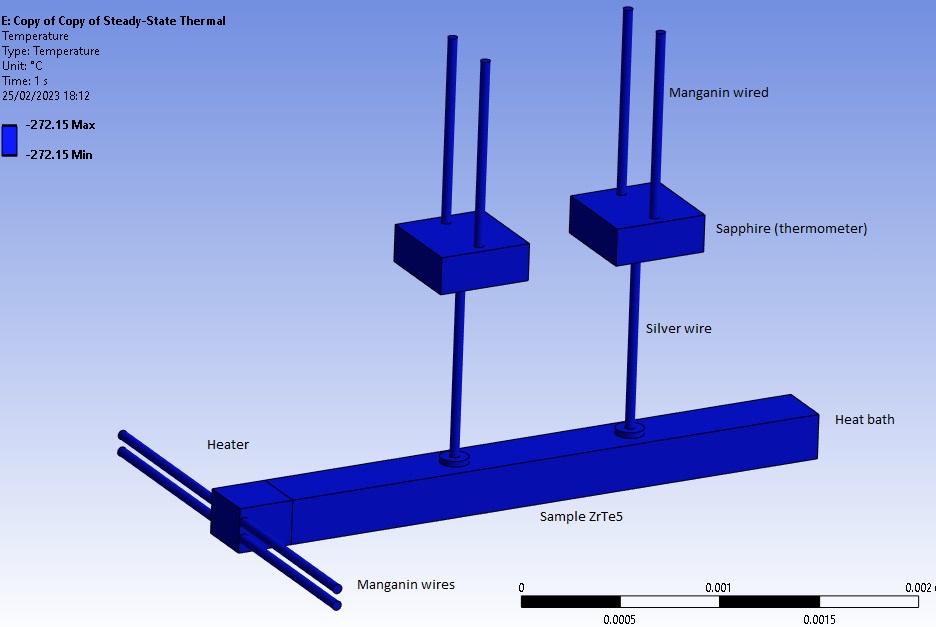}}
    \caption{Calculated temperature distribution across the sample, with no heat applied. Radiative corrections are calculated assuming the sample thermal bath being at 1K and surrounding chamber at 4K   
    }
    \label{no_gradient}
\end{figure*}

\begin{figure*}[th]
    \centerline{
    \includegraphics[width=0.6\textwidth]{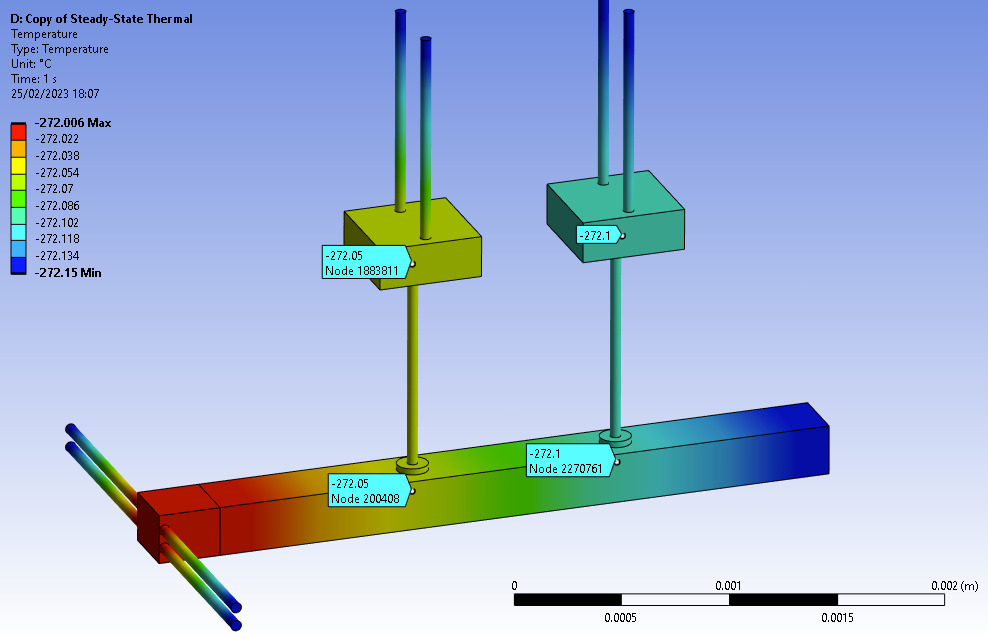}}
    \caption{ Calculated temperature distribution across the sample, with XX Watt power applied to the heater. Radiative corrections are calculated assuming the sample thermal bath being at 1K and surrounding chamber at 4K  
    }
    \label{gradient}
\end{figure*}

Simulations results shown on Fig.\ref{no_gradient} and Fig.\ref{gradient} confirm that in our design most heat flow occurs from the heater to the thermal bath via ZrTe5 sample and thus our measured sample thermal conductivity's are expected to closely reflect the actual values of thermal conductivity. Table 3 summarizes the heat flow in ZrTe5 for all three used heating powers.

\begin{table}
    \centering
    \begin{tabular}{|l|r|r|} \hline 
        Heatflow into ZrTe5&Heatflow via ZrTe5 to thermal bath&Difference \\ \hline 
        \textit{uW} &  \textit{uW}& \textit{\%} \\ \hline 
         0.025&  0.0246& 1.65\\ \hline 
         0.9&  0.0885& 1.65\\ \hline 
         2.5&  2.46& 1.65\\ \hline
    \end{tabular}
    \caption{Heat flow comparison for ZrTe5 sample in simulations}
    \label{tab:my_label}
\end{table}

Investigation of Fig.\ref{no_gradient} and Fig.\ref{gradient} confirms that in our design most heat generated by the heater is transported through the sample to the thermal bath and thus our measured sample thermal conductivity's are expected to closely reflect the actual values of thermal conductivity.

